\input phyzzx

\def\IR{{\hbox{{\rm I}\kern-.2em\hbox{\rm R}}}}
\def\IB{{\hbox{{\rm I}\kern-.2em\hbox{\rm B}}}}
\def\IN{{\hbox{{\rm I}\kern-.2em\hbox{\rm N}}}}
\def\IC{{\ \hbox{{\rm I}\kern-.6em\hbox{\bf C}}}}

\def\IZ{{\hbox{{\rm Z}\kern-.4em\hbox{\rm Z}}}}

\def\underarrow#1{\vbox{\ialign{##\crcr$\hfil\displaystyle
{#1}\hfil$\crcr\noalign{\kern1pt
\nointerlineskip}$\longrightarrow$\crcr}}}
%
\def\d{{\rm d}}

\def\hat{\widehat}
\def\gL{\Lambda}
\def\ga{\alpha}

\def\gd{\delta}

\def\p{\partial}
\def\gee{\epsilon}
\def\gl{\lambda}
\def\th{\theta}
\def\bx{{\bf x}}
\def\gO{\Omega}
\def\l{\left}
\def\r{\right}

\nopagenumbers
\line{\hfil CU-TP-696}
\line{\hfil IUHET-301}
\line{\hfil hep-th/9505103}
\vglue .4in
\centerline {\twelvebf  WZW model based on the extended de Sitter group}
\vskip .3in
\centerline{\it Hai Ren }
\vskip .1in
\centerline{Physics Department }
\centerline{Indiana University }
\centerline{Bloomington, IN 47405 }
\vskip .4in
\overfullrule=0pt
\centerline {\bf Abstract}
\medskip

We study the WZW model based on the centrally extended
2D de Sitter algebra. We obtain the spacetime metric and its
explicitly conformally flat expression. The symmetries of the
spacetime are found by identifying the Killing vectors with the group
generators. The energy-momentum tensor obtained from the
affine-Sugawara construction agrees with that from
the more conventional
approach. The exact center charge agrees to one-loop order with the
one-loop beta function equations. We have also
studied the representations of the
corresponding enveloping Virasoro algebra.

\vskip 1.1in
\noindent\footnote{}{\twelvepoint This work was supported in part by
the US Department of Energy }

\vfill
\eject

\baselineskip=19pt
\pagenumbers
\pageno=1

\overfullrule=0pt
\tolerance=5000
\overfullrule=0pt
\twelvepoint

\twelvepoint

\REF\list{O. Jofre and C. Nunez,
hep-th/9502150, and references therein;
\nextline
K. Sfetsos and A. A. Tseytlin, Nucl. Phys. B{\bf 427}, 
245 (1994), and references therein;
\nextline
I. Antoniadis and N. A. Obers, Nucl. Phys. B{\bf 423},
639 (1994).}

\REF\Witten{C. Nappi and E. Witten, Phys. Rev. Lett. {\bf 71},
3751 (1993).}

\REF\Jackiw{D. Cangemi and R. Jackiw, Phys. Rev. Lett. {\bf 69}, 233
(1992); Ann. Phys. (NY) {\bf 225}, 229 (1993);
D. Cangemi and G. Dunne, Phys. Rev. D{\bf 48}, 5721 (1993).}

\REF\book{See for example, M. B. Green, J. H. Schwarz and E. Witten,
{\it Superstring Theory}, Volume I (Cambridge University Press, 1987).}

\REF\fields{A. M. Polyakov and P. B. Wiegmann, Phys. Lett. B{\bf 141},
 223 (1984).}

\REF\KM{E. Witten, Comm. Math. Phys. {\bf 92}, 455 (1984).}

\REF\Halpern{M. B. Halpern and E. Kiritsis, Mod. Phys. Lett.
{\bf A4}, 1373 (1989).}

\REF\Kac{V. G. Kac and A. K. Raina, {\it Highest Weight Representations
of Infinite Dimensional Lie Algebras} (World Scientific, 1987).}

\REF\recent{J. M. Figueroa-O'Farrill and S. Stanciu,
Phys. Lett. B{\bf 327}, 40 (1994);
\nextline
N. Mohammedi, Phys. Lett. B{\bf 325}, 371 (1994);
Phys. Lett. B{\bf 331}, 93 (1994).}

\REF\olive{D. I. Olive, E. Rabinovici and A. Schwimmer,
Phys. Lett. B{\bf 321}, 361 (1994).}

\REF\HS{G. T. Horowitz and A. R. Steif, Phys. Rev. Lett. {\bf 64},
260 (1990); Phys. Rev. D{\bf 42}, 1950 (1990).}

\REF\duval{C. Duval, Z. Horv\'ath and P. A. Horv\'athy, Mod. 
Phys. Lett. {\bf A8}, 3749 (1993).}

\REF\ketov{S. V. Ketov, Nucl. Phys. B{\bf 294}, 813 (1987).}

\REF\CI{A list of these 14 curvature invariants is given in
A. Harvey, Class. Quant. Grav. {\bf 7}, 715 (1990);
K. Lake, J. Math. Phys. {\bf 34}, 5900 (1993).}

\REF\Farhi{S. M. Carroll, E. Farhi, A. H. Guth and K. D. Olum,
Phys. Rev. D{\bf 50}, 6190 (1994).}

\REF\differ{For discussions of obtaining group generators as
differential operators, see for example,
A. Wawrzynczyk, {\it Group Representations and Special Functions}
(Polish Scientific Publishers, 1984).}

\REF\DeWitt{B. S. DeWitt, in {\it Relativity, Groups and Topology}
(Gordon and Breach, Science Publishers, 1964);
I. Jack, D. R. T. Jones, N. Mohammedi and H. Osborn,
Nucl. Phys. B{\bf 332}, 359 (1990).}

\REF\measure{See, for instance, M. Hamermesh, {\it Group Theory
and Its Application to Physical Problems}
(Addison-Wesley, 1962).}

\REF\conformal{For a review, see
P. Ginsparg, {\it Applied conformal field theory},
in {\it Fields, Strings and Critical Phenomena}, edited by
E. Br\'ezin and J. Zinn-Justin (Les Houches lectures, 1988).}

\REF\geom{M. B. Halpern and J. P. Yamron, Nucl. Phys. B{\bf 332},
411 (1990).}

There has recently been much interest in WZW models based
on non-semi-simple groups [\list][\Witten].
The first of such models is constructed in
[\Witten], based on the centrally extended 2D Poincar\'e algebra,
which is used in analyzing the gauge theory of 2D
gravity models.
In this paper we consider the ``centrally
extended 2D de Sitter algebra'', which is also used in the gauge
formulation of lineal gravity models [\Jackiw].
The algebra has the following explicit description:
$$[J, P_a]=\epsilon_{ab}P_b \qquad [P_a, P_b]=\epsilon_{ab}(-\gL J + T)
\qquad [T,J]=[T,P_a]=0
	\eqn\zero$$
We will call the corresponding non-compact group $G$. The extended
Poincar\'e algebra is given by the above with $\gL = 0$.

In general, given a Lie  algebra with generators
$T^a$ (here $T^a= P_1, P_2, J, T$), and structure constants $f_d^{ab}$
(so $[T^a,T^b]=if_d^{ab}T^d$), if there is a bilinear form
$\Omega^{ab}$ in the
generators $T^a$, which is symmetric ($\Omega^{ab}=\Omega^{ba}$),
invariant
$$f^{ab}_d\Omega^{cd} + f^{ac}_d\Omega^{bd} = 0
	\eqn\inv $$
and non-degenerate (so that there is an inverse matrix $\Omega_{ab}$
obeying $\Omega_{ab}\Omega^{bc}=\delta_a^c$),
then the WZW action on the surface $\Sigma$
of a three-manifold $B$ is
$$ S(g) = {1\over 4\pi}\int_\Sigma \d^2\sigma ~\Omega^{ab} A_{a\ga}
A_b^\alpha + {i\over 12\pi}\int_B \d^3\sigma~
\epsilon_{\alpha\beta\gamma}A_a^\alpha A_b^\beta A_c^\gamma
\Omega^{cd}f_d^{ab} \eqn\wzw$$
where the $A_{a\alpha}$'s are defined via
the left invariant one-forms
 $ g^{-1}\partial_\alpha g = A_{a\alpha} T^a$.
Usually for semisimple groups one can choose the bilinear form
 $\Omega^{ab} = K^{ab}
\equiv f_d^{ac}f_c^{bd}$.
However, for
non-semi-simple groups this quadratic form is degenerate.
Nevertheless, the $G$ Lie algebra
does have another non-degenerate bilinear form [\Jackiw]
{\it i.e.}
$$\Omega^{ab}= k\pmatrix{1&0&0&0\cr
0&1&0&0\cr
0&0&{b\over 1-b\gL}&{1\over 1-b\gL}\cr
0&0&{1\over 1-b\gL}&{\gL\over 1-b\gL}\cr}
	\eqn\ndeg  $$
This metric on the Lie algebra has signature
$(+,+,+,-{\rm sign}(1-b\gL))$, and that will
therefore be the signature of the resulting
space-time metric.

In order to write \wzw\ explicitly we need to find the $A_a$'s. To
this purpose we use the following parametrization of the group
manifold:
$$g= e^{a_1P_1 + a_2P_2}e^{uJ+vT}   \eqn\group$$
from which we obtain
$$\eqalign{A_{a\alpha} A_b^\alpha\Omega^{ab}&= \partial_\alpha
a_k\partial^{\alpha}a_k + {1\over 1 - b\gL}(
b\partial_\alpha u\partial^\alpha u +
2 \partial_{\alpha}v\partial^{\alpha}u +
\gL\p_\ga v\p^\ga v) \cr
\qquad &+
\partial^{\alpha}u
{2(1 - \cosh q)
\over q^2}\epsilon_{ij}
a_i\partial_\alpha a_j
 + \gL
{2(\cosh q - 1) - q^2\over q^4}(\gee_{ij}a_i\p_\ga a_j)^2 }
	\eqn\kin$$

$$\epsilon_{\alpha \beta \gamma} A_a^\alpha A_b^\beta A_c^\gamma
\Omega^{cd}f_d^{ab} =2\epsilon_{\alpha \beta \gamma}\times
3\partial^{\alpha}[u{\sinh q\over q}
\partial^\beta a_1\partial^\gamma a_2]
	\eqn\wzterm$$
where
$ q \equiv \sqrt{\gL(a_1^2 + a_2^2)}$.
Thus we are still able to reduce the Wess-Zumino term to a surface
term without introducing any singularities.
By using the polar coordinates
$$ a_1 \equiv r\cos\th  \qquad\qquad   a_2 \equiv r\sin\th
	\eqn\eq   $$
and identifying the resulting action with the non-linear
$\sigma$-model action of the
form
$$S = {k\over 4\pi}
\int \d^2\sigma (G_{ab}\partial_\alpha X^a{\partial^\alpha} X^b
+iB_{ab}\epsilon_{\alpha\beta}\partial^\alpha X^a\partial^\beta X^b)
	\eqn\six  $$
where $X^a=(r, \th, u, v)$,
one can
read off the background space-time metric and antisymmetric tensor field.
The space-time geometry  is described by the  metric 
$$ ds^2= dr^2 + 2{\cosh\gl r - 1\over \gl^2}d\th^2
- 2{\cosh\gl r - 1\over \gl^2}du d\th + {1\over 1 - b\gL}(bdu^2
+ 2du dv + \gL dv^2)
\eqn\name$$
and
$$ B_{12} = u{\sinh\gl r\over \gl}
	\eqn\eq  $$
where $\gl \equiv \sqrt\gL$ ,
and the dilaton is constant because of
the homogeneity of the group manifold.
In terms of $(r, \th, u, v)$, the left invariant currents
are
$$\eqalign{A_{1\ga} &= \cos(u-\th)\p_\ga r
+ \sin(u-\th){\sinh \gl r\over \gl}\p_\ga \th   \cr
A_{2\ga} &= -\sin(u-\th)\p_\ga r
+ \cos(u-\th){\sinh \gl r\over \gl}\p_\ga \th   \cr
 A_{3\ga} &= \p_\ga u + (\cosh \gl r - 1)\p_\ga\th   \cr
 A_{4\ga} &= \p_\ga v - {1\over \gl^2} (\cosh \gl r - 1)\p_\ga\th  }
	\eqn\pole    $$
Introducing
$$ A_{a\ga} \equiv A_{ai}\p_\ga X^i \equiv
(T^{-1})_{ia}\p_\ga X^i
	\eqn\ts   $$
we have
$$ \p_i(T^{-1})_{ja} - \p_j(T^{-1})_{ia} = i f^{bc}_a(T^{-1})_{ib}
(T^{-1})_{jc}
	\eqn\cartan    $$
These are the Maurer-Cartan equations satisfied by the left invariant
currents. 
The classical equations of motion are given by,
in terms of the light-cone coordinates
$x_\pm = {1\over \sqrt{2}}(\tau \pm \sigma)$,
$$ \p_- A_{a+} = 0
	\eqn\eq   $$
Thus the $A_{a+}$'s are functions of $x_+$ only.

To check that this model is conformally invariant,
we first look at the one loop beta function
equations [\book]
$$\eqalign{&R_{ab} + {1\over 4}H_{ab}^2 + \nabla_a \nabla_b\phi=0\cr
 & \nabla^c H_{cab} + \nabla^c\phi
H_{cab} = 0\cr & R + {1\over {12}}H^2 + 2\nabla^2\phi +
(\nabla\phi)^2 - \Lambda_1 =0\cr}
	\eqn\uffa$$
where $H_{cab} = \nabla_{[c}B_{ab] }$ and
$H^2_{ab}=H_{acd}H_b^{\ cd}$, $H^2=H_{abc}H^{abc}$,
and $\Lambda_1=2k(c-4)/3 $.
One finds that the non-zero components of $R_{ab}$ are
$$ R_{11} = {1\over 2}\gl^2 \qquad R_{22} = \cosh\gl r - 1 \qquad
R_{23} = -{1\over 2}(\cosh\gl r - 1) \qquad
R_{33} = -{1\over 2}
	\eqn\eq  $$
and as the only non-zero component of $B$ is
$B_{12}=u{\sinh\gl r\over \gl}$,
the only non-zero component of $H$ is $H_{123}={\sinh\gl r\over \gl}$.
Also we have $R=3\gL/2$ and $H^2 = -6\gL $.
Putting these pieces together, one verifies equations \uffa\
with $\Lambda_1 = \gL$ and $ c = 4 + {3\gL\over 2k} $.

This can also be studied nonperturbatively by generalizing
the Sugawara construction to non-semi-simple algebras.
The Lie algebra
of Eq.\zero\ admits a $2\times 2$ representation
$$ P_1 = {\gl\over 2}
\pmatrix{0&1\cr
1&0\cr} \quad
P_2 = {\gl\over 2}
\pmatrix{0&-i\cr
i&0\cr} \quad
J =
\pmatrix{b'-{1\over 2}i&0\cr
0&b'+{1\over 2}i\cr} \quad
T = b'\pmatrix{\gl^2&0\cr
0&\gl^2\cr}
	\eqn\twod   $$
One finds that (here $t^a = P_1, P_2, J, T$)
$$ {\rm Tr}\ t^a t^b = {\gL\over 2k}\gO^{ab}
	\eqn\eq   $$
provided $4b'^2 = 1/(1 - b\gL)$.
Thus the Polyakov-Wiegmann composition rule still holds, 
and one can obtain the Kac-Moody symmetry of the
model [\fields][\KM]. Introducing
$$ J^a(x_+) \equiv 2\gO^{ab}A_{b+}(x_+)
	\eqn\eq   $$
we have the canonical commutators
$$ [J^a(x_+), J^b(y_+)] = -2\pi\gd(x_+ - y_+)f^{ab}_cJ^c(x_+)
 - 4\pi i\gd'(x_+ - y_+)\gO^{ab}
	\eqn\eq   $$
These currents correspond
to fermionic currents in a free fermion theory 
under the non-abelian bosonization.
Introducing $z = e^{ix_+}$ (hence $|z| = 1$)
and expanding $J^a(z)$ in Laurent
series
$$ J^a(z) = \sum_{n = -\infty}^\infty J^a_n z^{-n-1}  \qquad\quad
J^a_n = \oint {dz\over 2\pi i}z^n J^a(z)
	\eqn\eq   $$
one obtains
$$ [J^a_m, J^b_n] = if^{ab}_cJ^c_{m+n} + 2\gO^{ab}m\gd_{m+n,0}
	\eqn\modealgebra   $$
which can be viewed as a representation of the Kac-Moody algebra
in terms of the currents.
Following [\Halpern] we take the moments of the
energy-momentum tensor to be
$$ L_m = L_{ab}\sum_k:J^a_{m+k}J^b_{-k}:\
\equiv L_{ab}\l[\sum_{k\le -m/2}J^a_{m+k}J^b_{-k} +
\sum_{k > -m/2}J^b_{-k}J^a_{m+k}\r]
	\eqn\eq   $$
where $L_{ab} = L_{ba}$. Namely
$T(z) = \sum_{n = -\infty}^\infty L_n z^{-n-2}$
is the energy-momentum tensor operator. 
Our normal ordering convention is the same as
that used in [\Kac]. 
Hermiticity of these generators implies
$ J^{a\dagger}_n = J^a_{-n}$ and
$ L^\dagger_n = L_{-n}$.
Using Eq.\modealgebra,
we calculate
$$ \eqalign{[L_m, J^a_n] &= L_{cd}\sum_k
\l[ J^c_{m+k}J^d_{-k}, J^a_n\r] \psi(\gee k) \cr
&= -n\l(4L_{cd}\gO^{ac} + L_{bc}f^{ac}_ef^{eb}_d\r)J^d_{m+n}
 + iL_{cd}f^{ac}_e\l\{\sum_k:J^d_{m+k}J^e_{-k}:\   \r.\cr
&\qquad + \l.\sum_k:J^e_{m+k}J^d_{-k}:\  + \
2\gO^{de}\gd_{m+n,0}\l[\sum_{k>-m}(k+m) + \sum_{k>0}k\r]
\psi(\gee k) \r\}   }
	\eqn\eq  $$
where the symmetric cut-off function [\Kac]
$$ \psi(x) = \cases{ 1   &$\qquad |x| \le 1$  \cr
0    &$\qquad |x| > 1$ }
	\eqn\eq   $$
is to regularize possible ambiguities under shifting indices.
We have suppressed the function $\psi(x)$ for those terms that do not
have such ambiguities.
Requiring that the currents be primary fields of the
energy-momentum tensor with conformal weight one
$$ [L_m, J^a_n] = -nJ^a_{m+n}
	\eqn\pime   $$
one finds [\Halpern][\recent] that $L_{ab}$ is invariant and that
$$ L^{-1}_{ab} = 4\gO^{ab} + K^{ab}
	\eqn\eq  $$
Next one can use Eq.\pime\ to verify the Virasoro algebra
$$\eqalign{[L_m, L_n] &= L_{cd}\sum_k[L_m, J^c_{n+k}J^d_{-k}]
\psi(\gee k)  \cr
&= (m-n)L_{m+n} + {1\over 3}(m^3 - m) L_{cd}\gO^{cd}\gd_{m+n,0} }
	\eqn\Vira   $$
Therefore $ c = 4L_{ab}\gO^{ab}$.
In our case, the inverse inertia tensor is given by
$$L_{ab} =
{1\over 2(2k-\gL)}\pmatrix{1&0&0&0\cr
0&1&0&0\cr
0&0&-\gL&1\cr
0&0&1&-b - {1 - b\gL\over 2k}\cr}
	\eqn\tensor$$
This is the expected result since our model is equivalent 
to an $SU(1,1)\otimes \IR $ WZW model. 
The center charge is
$$ c = 4 + {3\gL\over 2k-\gL} = {6k\over 2k-\gL} + 1
 = 4 + {3\gL\over 2k} + O\l({1\over k^2}\r)
	\eqn\charge  $$
where the contribution $6k/(2k-\gL)$
is from the $SU(1,1)$ factor. A similar
analysis has been carried out for a larger class of non-semi-simple
groups [\olive].

Conformally invariant theories with plane wave metrics were studied
in [\HS]. It was shown that the beta function equations are
satisfied to all loop orders due to the vanishing of
curvature invariants. The study was extended to cases where
the metrics are of the form of
Brinkmann's generalized plane-fronted waves with
parallel rays 
in Ref.[\duval]. This method does not apply here, since all curvature 
invariants do not
vanish. (And hence that rank two tensors such
as $R^2R_{ab}$ do not vanish.)

It remains interesting to study the construction of these
curvature invariants. 
One finds that the metric of Eq.\name\ is conformally flat, and that
$$  \nabla_f R_{abcd} = 0\qquad\qquad\qquad \nabla_f H_{abc} = 0
	\eqn\condd   $$
Hence the beta functions must all be covariantly constant
to all loop orders.
In particular, $\beta^\Phi$ must be a constant to all
orders in $(1/k)$.
It was conjectured that the conditions Eqs.\uffa\
and \condd\ 
imply that the Wess-Zumino term is never renormalized to all
loop orders [\ketov].
Now let us consider a curvature invariant of order $k$
$$ Q = Q(g_{ab}, R_{abcd}, \ldots,
\nabla_{f_1}\ldots\nabla_{f_k}R_{abcd})
	\eqn\eq   $$
We see that only $0$-th order invariants can be constructed. Moreover,
the $14$ $0$-th order curvature invariants of Eq.\name\
are all constants, due to
$$ \p_f Q = \nabla_f Q = 0
	\eqn\eq   $$
Explicitly, these invariants are the $10$ vanishing quantities
constructed with the use of the Weyl tensor
and the $4$ invariants formed out of the Ricci tensor [\CI]
$$ Q_n \equiv
{\rm Tr}\left(R_a^b\right)^n = {3\gL^n\over 2^n}\qquad\qquad
\qquad n = 1, 2, 3, 4
	\eqn\eq  $$

The model describes
a homogeneous space-time; the left and right action of $G$ on itself
gives rise to a seven dimensional symmetry group of the space-time,
as the left and right actions of the
central generator $T$ coincide.
By a shift in $v$ and then 
a rotation with angle $u$ in the $(a_1, a_2)$ plane,
we can re-write 
Eq.\name\ as $ds^2 = d\hat s^2 + dv^2$ with
$$ d\hat s^2 = {4\over \gL}( dr^2 + \sinh^2 r\
d\th^2 - \cosh^2 r\ du^2 )
	\eqn\dS  $$
being a metric for the 3D de Sitter space,
which can be realized as an embedding in the 4D flat
space
$$ ds^2 = -dt^2 + dx^2 + dy^2 -dw^2
	\eqn\flat  $$
with the transformations [\Farhi]
$$ t = {2\over \gl}\sin u\cosh r  \qquad
 x = {2\over \gl}\cos \th \sinh r  \qquad
 y = {2\over \gl}\sin \th \sinh r  \qquad
 w = {2\over \gl}\cos u \cosh r
 	\eqn\eq  $$
We see that the seven symmetries of Eq.\name\
are associated with the six symmetries of the 3D de Sitter
space and a translation in the flat dimension.
Using a different set of embedding
coordinates
$$ t = \cos\chi  \qquad
 x = \sin\chi\sinh\eta\cos\phi  \qquad
 y = \sin\chi\sinh\eta\sin\phi  \qquad
 w = \sin\chi\cosh\eta
	\eqn\eq  $$
one can bring the metric
$$ ds^2 = dv^2 + dr^2 + \sinh^2 r\
d\th^2 - \cosh^2 r\ du^2
	\eqn\cflat   $$
to the Robertson-Walker form
$$ ds^2 = dv^2 - d\chi^2 + \sin^2\chi\l(d\eta^2
+ \sinh^2\eta\ d\phi^2\r)
	\eqn\eq   $$
whose explicitly conformally flat expression is known.
We thus find
$$ ds^2 = \cos^2{1\over 2}(v+\chi)\cos^2{1\over 2}(v-\chi)
\l[ 4 d\alpha d\beta + (\alpha - \beta)^2
( d\eta^2 + \sinh^2\eta\ d\phi^2)\r]
	\eqn\flattt   $$
where
$$ \alpha = \tan{1\over 2}(v + \chi)  \qquad\quad
\beta = \tan{1\over 2}(v - \chi)
	\eqn\flatttt   $$
The spacetime inside the bracket in Eq.\flattt\ is manifestly
flat. The result can be viewed as a generalization
of the fact that the exact plane wave studied in [\Witten]
$$ ds^2 = d\bx^2 -2du\l[dv + {1\over 2}\gee_{ij}x^jdx^i\r] 
	+ bdu^2
	\eqn\eq    $$
is conformally flat with conformal factor $\cos^{-2}(u/2)$.
The above space-time has non-vanishing Ricci tensor, a
covariantly constant null vector and vanishing
curvature invariants. It also has seven symmetries.

The explicit forms of the
symmetry of the metric Eq.\name\ can be found
as follows. The group generators, when acting on group
parameter space, can be represented as
differential operators [\differ].
With the parametrization of Eq.\group, we have
$$\eqalign{ P_1 &= \cos(u-\th) {\p\over \p r}
+ {\gl\sin(u - \th)\over \sinh\gl r}{\p\over \p \th}
- \sin(u - \th)\tanh{\gl\over 2}r\l(
\gl{\p\over \p u} - {1\over \gl}{\p\over \p v}\r)   \cr
 P_2 &= -\sin(u-\th) {\p\over \p r}
+ {\gl\cos(u - \th)\over \sinh\gl r}{\p\over \p \th}
- \cos(u - \th)\tanh{\gl\over 2}r\l(
\gl{\p\over \p u} - {1\over \gl}{\p\over \p v}\r)   \cr
 J &= {\p\over \p u}   \cr
 T &= {\p\over \p v}   }
	\eqn\intrgen  $$
They correspond to the right action of the group,
{\it i.e.},
with infinitesimal elements acting from 
the right.
We can write the above $T^a$'s (here $T^a = P_1, P_2, J, T$) as
$$ T^a = (T^a)^i {\p\over \p X^i}
= T^{ai}{\p\over\p X^i}
	\eqn\gene   $$
where the $T^{ai}$'s are introduced in Eq.\ts. Indeed, from the
structure equation Eq.\cartan\ and $\p(T T^{-1}) = 0$ one
obtains
$$ T^{ai}\p_i T^{bj} - T^{bi}\p_i T^{aj} = i f^{ab}_c T^{cj}
	\eqn\comm   $$
which are the desired group algebras.

It follows that
the $T^a$'s are Killing vectors of the metric [\DeWitt],
which can be expressed as
$$ G^{ij} = \gO_{ab}T^{ai}T^{bj}
	\eqn\Gij   $$
In fact, by
using the commutators Eq.\comm\ and the invariant
condition Eq.\inv, one can verify
$$ T^{ak}\p_k G^{ij} - G^{kj}\p_k T^{ai}
- G^{ik}\p_k T^{aj} = 0
	\eqn\killing   $$
which are equivalent to the Killing equations.
The group invariant measure 
is found to be proportional to
the volume element of the metric Eq.\name:
$$ \l|{\rm det} T^{ai}\r|^{-1} = \l|{\rm det} \gO_{ab}\r|^{1/2}
\sqrt G = {\sinh \gl r\over \gl}
	\eqn\eq   $$
Furthermore, the quadratic Casimir operator is
equal to the Laplacian associated with
the metric:
$$ \Delta \equiv
\gO_{ab}T^aT^b = {1\over\sqrt G}\p_i\l(\sqrt G G^{ij}\p_j\r)
	\eqn\eq   $$
To see this, we use the formula, which is true for an arbitrary
non-singular matrix $X$,
$$ {\p {\rm det}X\over {\rm det}X}
= {\rm Tr}\l( X^{-1}\p X\r)
	\eqn\eq   $$
and the commutation relations Eq.\comm, to obtain
$$ \gO_{ab}T^aT^b - {1\over\sqrt G}\p_i\l(\sqrt G G^{ij}\p_j\r)
= - \gO_{ab}f^{ac}_c T^{bj}\p_j
	\eqn\eq   $$
But this vanishes due to the invariant condition Eq.\inv:
$$ - \gO_{ab}f^{ac}_c T^{bj}\p_j
=  \gO_{ac}f^{ac}_b T^{bj}\p_j = 0
	\eqn\eq   $$
The group representation functions,
which we denote as $|lmp>$,
are simultaneous eigenfunctions
of $\Delta, J$ and $T$. They
form the regular representation of the group.

The remaining three Killing vectors correspond to the left action
of the group. The corresponding group generators
are given by
$$\eqalign{\overline P_1 &= \cos\th {\p\over \p r} - \gl\sin\th
\coth\gl r{\p\over \p \th} - \sin\th\tanh{\gl\over 2}r\l(
\gl{\p\over \p u} - {1\over \gl}{\p\over \p v}\r)   \cr
\overline P_2 &= \sin\th {\p\over \p r} + \gl\cos\th
\coth\gl r{\p\over \p \th} + \cos\th\tanh{\gl\over 2}r\l(
\gl{\p\over \p u} - {1\over \gl}{\p\over \p v}\r)   \cr
\overline J &= {\p\over \p \th} + {\p\over \p u}   \cr
\overline T &= {\p\over \p v}   }
	\eqn\generator  $$
From the above one obtains the left invariant measure and the
Casimir operator. In this case they are the same as the right
invariant ones.
The two sets of generators are independent ($T$ and $\overline T$
are the same). Together they form the anti-de Sitter algebra
$so(2,2)$ plus a center element.
Since the metrics constructed
via the left or right invariant
one-forms are the same, the left and right invariant measures will
always be the same whenever such metrics can be 
constructed [\measure].
Note that at any given point, we can always find four Killing
vectors $\l( {\partial\over \partial r}, {\partial\over \partial\th},
{\partial\over \partial u}, {\partial\over \partial v} \r) $,
as required by the homogeneity of the spacetime.

In the following we consider highest weight representations of the
enveloping Virasoro algebra of Eq.\Vira.
The states $|lmp>$ provide representations for the
zero mode generators $J^a_0$. 
Assuming there exists a highest weight state
$|lmp, N>$, where $N$ denotes additional degrees of freedom,
satisfying [\conformal]
$$ J^a_n |lmp, N>\ = 0 \qquad  L_n |lmp, N>\ = 0 \qquad\quad
(n > 0)
	\eqn\eq   $$
The Virasoro operator $L_0$, when acting on this
highest weight state,
can be represented as
$$ L_0 = L_{ab}J^a_0J^b_0 = L_{ab}T^aT^b
	\eqn\eq  $$
Since $L_{ab}$ is invariant, $L_0$ can be identified with the
Laplace operator associated with the metric
$$ \widetilde G^{ij} = L_{ab}T^{ai}T^{bj}
	\eqn\eq   $$
Note that $\widetilde G^{ij}$ is proportional to $G^{ij}$ up
to a redefinition of $b$. We then obtain
$$ L_0 = \overline L_0 = {k\over 2(2k - \gL)}\Delta -
{1 - b\gL\over 4k(2k - \gL)}{\partial^2\over \partial v^2}
=  {k\over 2(2k - \gL)}\overline\Delta -
{1 - b\gL\over 4k(2k - \gL)}{\partial^2\over \partial v^2}
	\eqn\eq   $$
This gives the conformal weight $h_{lmp}$
of the highest weight state in terms
of the eigenvalues of $\Delta$ and $T$.
In the geometric formulation of the general affine-Virasoro
construction, the center charge is given by [\geom]
$$ c = {\rm dim}\ G + 4\widetilde R
	\eqn\eq   $$
where $\widetilde R$ is the curvature scalar associated with the
metric $\widetilde G^{ij}$. One easily verifies that this reproduces
our previous result for the center charge.

Descendant states are formed by the action of a series of
$L_{-n}$ ($ n > 0$)
on $|lmp, N>$. 
A linear combination of states that vanishes is known as a null
state. The representation of the Virasoro algebra is
constructed from the Verma module consisting of the highest
weight state and the set of states
$$ L_{-k_1}L_{-k_2}\ldots L_{-k_n}|lmp, N>  \qquad\quad
1\le k_1\le k_2 \le \ldots \le k_n 
	\eqn\eq   $$
by removing all null states and their descendants.
Unitarity of the representation requires the non-existence
of negative norm states.
(If $O^\dagger O|\alpha> = - |\alpha>$, then $O|\alpha>$ is a 
``negative norm'' state.)  
This in turn requires that the Kac
determinant be non-negative, which results in the conditions
$c \ge 1$ and $h_{lmp} \ge 0$, in addition to the discrete
series with $0 < c < 1$.
If our parameters are such that
$c > 1$, so that there are no null descendant states except at
$h_{lmp} = 0$, we can write down the Virasoro characters as
$$  \chi_{lmp}(q, \th_1, \th_2) = q^{-c/24}
{\rm Tr}_{lmp}q^{L_0}e^{i\th_1 J + i\th_2 T}
= { q^{-(c/24)+h_{lmp}}e^{i(m+p)\th_1 + i\gL p\th_2} \over
\prod^\infty_{n=1}(1-q^n) }
	\eqn\eq   $$
for $h_{lmp} \ne 0$, and
$$  \chi_{lmp}(q, \th_1, \th_2)
= { q^{-c/24}e^{i(m+p)\th_1 + i\gL p\th_2} (1 - q) \over
\prod^\infty_{n=1}(1-q^n) }
	\eqn\eq   $$
for $h_{lmp} = 0$.

In conclusion,
we have studied the WZW model based on the ``centrally extended
2D de Sitter algebra''. We found the explicitly conformally
flat expression of the spacetime metric, described by 
Eqs.\flattt\ and \flatttt.
The energy-momentum tensor obtained from the
affine-Sugawara construction agrees with that from
the more conventional
approach. The exact center charge agrees to one-loop order with the
one-loop beta function equations. We have also
studied the representations of the
corresponding Virasoro algebra.

\ack

	I thank Alan Kosteleck\'y,
Kimyeong Lee and  Erick J. Weinberg
for helpful discussions. I also thank M. B. Halpern, 
A. Kehagias, O. Jofre,
C. Nunez, K. Sfetsos and A. A. Tseytlin for their comments.
This work was started when I was
at the Physics Department of Columbia University.

\refout

\end